\documentclass[fleqn,usenatbib]{mnras}
\usepackage[T1]{fontenc}
\usepackage{graphicx}
\usepackage{amsmath}
\usepackage{amssymb}
\usepackage{xspace}
\usepackage{newtxtext,newtxmath}
\usepackage{multirow}
\usepackage{lipsum}
\newcommand{\desyo}{DES-Y1\xspace}

\defcitealias{2206.15394}{J22}

\defcitealias{2018MNRAS.477.1822M}{M18}

\newcommand{\pcal}{\boldsymbol{\nu}}
\newcommand{\pncal}{\boldsymbol{\Omega}}

\newcommand{\cN}{\mathcal{N}}
\newcommand{\Om}{\Omega_{\rm m}}
\newcommand{\sig}{\sigma_{\rm 8}}

\newcommand{\Ob}{\Omega_{\rm b}}
\newcommand{\ns}{n_{\rm s}}

\newcommand{\nv}{\hat{\bf n}}

\newcommand{\lcdm}{$\Lambda$CDM\xspace}

\newcommand{\ccl}{\texttt{CCL}\xspace}

\newcommand{\mcal}{\textsc{Metacalibration}\xspace}

\DeclareRobustCommand{\VAN}[3]{#2}
\let\VANthebibliography\thebibliography
\def\thebibliography{\DeclareRobustCommand{\VAN}[3]{##3}\VANthebibliography}

\title[Analytical marginalisation over photo-$z$ uncertainties]{Analytical marginalisation over photometric redshift uncertainties in cosmic shear analyses}

\author[Ruiz-Zapatero et al.]{Jaime Ruiz-Zapatero$^{1}$ \thanks{E-mail: jaime.ruiz-zapatero@physics.ox.ac.uk},
Boryana Hadzhiyska$^{2, 3}$,
David Alonso$^{1}$,
Pedro G. Ferreira$^{1}$,
Carlos Garc\'ia-Garc\'ia$^{1}$
\newauthor
~and  Arrykrishna Mootoovaloo$^{1}$
\\
$^{1}$Department of Physics, University of Oxford, Denys Wilkinson Building, Keble Road, Oxford OX1 3RH, UK\\
$^{2}$Miller Institute for Basic Research in Science, University of California, Berkeley, CA, 94720, USA. \\
$^{3}$Physics Division, Lawrence Berkeley National Laboratory, Berkeley, CA 94720.
}

\date{Accepted XXX. Received YYY; in original form ZZZ}

\pubyear{2022}

\begin{document}
\label{firstpage}
\pagerange{\pageref{firstpage}--\pageref{lastpage}}
\maketitle

\begin{abstract}
  As the statistical power of imaging surveys grows, it is crucial to account for all systematic uncertainties. This is normally done by constructing a model of these uncertainties and then marginalizing over the additional model parameters.
 The resulting  high dimensionality of the total parameter spaces makes inferring the cosmological parameters significantly more costly using traditional Monte-Carlo sampling methods. A particularly relevant example is the redshift distribution, $p(z)$, of the source samples, which may require tens of parameters to describe fully. However, relatively tight priors can be usually placed on these parameters through calibration of the associated systematics. In this paper we show, quantitatively, that a  linearisation of the theoretical prediction with respect to these calibratable systematic parameters allows us to analytically marginalise over these extra parameters, leading to a factor $\sim30$ reduction in the time needed for parameter inference, while accurately recovering the same posterior distributions for the cosmological parameters that would be obtained through a full numerical marginalisation over 160 $p(z)$ parameters. We demonstrate that this is feasible not only with current data and current achievable calibration priors but also for future Stage-IV datasets.
\end{abstract}


\section{Introduction}\label{sec:intro}
  In recent years unprecedentedly precise observations in cosmology have uncovered a number of tensions between datasets that may constitute both tantalising hints of new physics or a manifestation of a lack of control over theoretical systematics \citep{K1K,2022ApJ...934L...7R}.

  At its simplest, the current cosmological paradigm, the $\Lambda$ (denoting the cosmological constant) cold dark matter model ($\Lambda$CDM), can be described by only five parameters: $\Omega_m$, $\Omega_b$, $A_s$, $n_s$ and $h$ (see e.g. \citet{LCDM} for a detailed review). However, in order to relate the theoretical predictions of this model to actual physical observables, it is necessary to extend it. Phenomenological models that describe the astrophysical systems that form the basis of our observations, as well as observational sources of systematic uncertainty, are then appended to the core $\Lambda$CDM model. In the presence of large statistical uncertainties, these models may consist of simple relationships in terms of a handful of parameters. However, more precise data requires an equally precise characterisation of these relationships, which leads to an increase in the complexity of the model. Thus, the number of parameters associated with these bridging models, colloquially referred to as ``nuisance'' parameters, has steadily grown over the years.

  The term ``nuisance'' is accurate when describing these parameters. Not only are they generally uninteresting by comparison with the fundamental cosmological parameters we aim to constraint, but the increase in parameter dimensionality of the model makes exploring their posterior distribution significantly more computationally costly. Standard Markov Chain Monte-Carlo (MCMC), and other rejection-based sampling methods  \citep[among  others]{Metropolis,  emcee, NS_Handley} suffer from the so-called ``curse of dimensionality'', whereby the acceptance rate of new samples decreases sharply with the number of parameters (exponentially in the worst cases). 
  
  Nuisance parameters can be divided into two groups based on their prior distributions: calibratable and non-calibratable parameters. The non-calibratable parameters can only be constrained by the data and, as such,  typically have largely non-constraining priors. On the other hand, we can place tighter priors on the calibratable parameters, either by accurately characterising the instrument measurements or by using independent external observations. In the case of cosmic shear analyses, the impact of galaxy intrinsic alignments \citep{astro-ph/0406275} is a standard example of a non-calibratable systematic. On the calibratable side, the two best examples are multiplicative shape measurement systematics, and the uncertainties in the redshift distribution of the target source galaxies \citep{2018MNRAS.478..592H, 2019MNRAS.483.2801S, 2020A&A...633A..69H,  Stolzner_Nz, Tianqing_Nz}.
  
  Of these calibratable systematics the dominant source of uncertainty in photometric surveys is the accuracy of redshift distributions, which are known to strongly affect the accuracy of cosmological constraints. The vital quantity to determine is the redshift distribution of each tomographic sample of galaxies, $p(z)$. The fact that the uncertainties in $p(z)$ can be calibrated with external spectroscopic data (e.g. via direct calibration, \citep{NIR_Nz1, NIR_Nz2}, clustering redshifts \citep{cross_corr_Nz1, cross_corr_Nz4, cross_corr_Nz2, cross_corr_Nz3}, and shear ratios \citep{DESY1_shear_ratios, DESY3_shear_ratios}), enables us to place relatively strong priors on the redshift distribution, which in turn makes it possible to use approximate methods to efficiently marginalise over these uncertainties.

  Analytical marginalisation schemes for photometric redshift uncertainties have already been proposed in the literature. In \citet{Stolzner_Nz} an analytic marginalisation scheme for photometric redshift uncertainties was proposed based on Gaussian mixture models and applied to the analysis of KV450 data \citep{KV450}. Alternatively, in \citet{Tianqing_Nz} a resampling approach to marginalize over these uncertainties was proposed and applied to the analysis HSC data. Here, we will explore the method initially proposed in \citet{Boryana_Nz_2020}, further exploited in \citet{2023JCAP...01..025G}, and recently characterised in the context of the Laplace approximation in \citet{FastBias}. The method is based on linearising the dependence of the theoretical prediction with respect to the parameters defining the redshift distribution around their calibration priors. This then allows one to analytically marginalise over these parameters by modifying the covariance matrix of the data, effectively assigning higher variance (as allowed by the calibration prior) to the data modes most sensitive to variations in the $p(z)$.
  
  The goal of this paper is to exhaustively validate this approximate marginalisation scheme in the context of cosmic shear analyses. We will do so by proving that we are able to obtain the same constraints on cosmological parameters using this scheme, as well as employing brute-force methods that sample the full parameter space exactly. We will show this for both simple parametrisations of the $p(z)$ uncertainties, in terms of shifts to the mean of the distribution, as well as using completely general ``non-parametric'' models that treat the amplitude of the $p(z)$ in narrowly-spaced intervals of $z$ as calibratable variables, leading to a model with more than $\sim100$ nuisance parameters. In order to numerically marginalize over such large parameter spaces we develop an auto-differentiable code to obtain theoretical predictions for the cosmic shear observables. This allows us to employ gradient based sampling algorithms, such as Hamiltonian Monte Carlo, to beat the aforementioned curse of dimensionality. Finally, we will show that the method is valid not only for current data, but also for futuristic Stage-IV surveys, where photometric redshift uncertainties will likely make up a large fraction of the total error budget. Interestingly, our analysis will show that, in the context of cosmic shear data, relatively inexpensive parametrisations of photometric redshift uncertainties based on one free parameter per redshift bin (e.g. mean shifts, or ranked discrete realisations \citep{2022MNRAS.511.2170C}), return effectively the same posterior distribution on cosmological parameters as the most general non-parametric models.

  This paper is structured as follows. In Section \ref{sec:methods} we describe the methods used in this work including the theory behind weak lensing observables, the calibration of redshift distributions, and the mathematics of analytical marginalisation via first-order expansion. Section \ref{sec:data} presents the Dark Energy Survey data used to produce realistic source redshift distributions and their associated uncertainties, as well as the models used to simulate future datasets. In Section \ref{sec:likelihood} we describe the likelihood used to analyse these data, as well as the different parametrisations used to describe $p(z)$ uncertainties. Section \ref{sec:results} presents our results, quantifying the performance of analytical marginalisation methods. Finally, we present our conclusions in Section \ref{sec:conc}.

\section{Methods}\label{sec:methods}
  \subsection{Cosmic shear power spectra}\label{ssec:methods.theo}
    It is now commonplace to carry out the analysis of galaxy weak lensing data tomographically. The full sample is split into redshift bins and the two-point correlation functions of all pairs of bins are measured and compared with their theoretical prediction. Let $\gamma_\alpha(\nv)$ be a map of the spin-2 lensing shear field inferred from the sources in the $\alpha$-th redshift bin. Its relation with the three-dimensional matter overdensity $\delta_m({\bf x})$ is  \citep{2001PhR...340..291B,2017arXiv170609359K}
    \begin{align}
      \gamma_\alpha(\nv)=\int _0^{\chi_H}d\chi\,q_\alpha(\chi)\,\left[-\chi^{-2}\eth\eth\nabla^{-2}\delta_m(\chi\nv,z)\right],\label{eq:deltagamma} 
    \end{align}
    where $\nv$ is the sky direction, $\chi$ is the comoving radial distance at redshift $z$, $\chi_H$ is the distance to the horizon, $q^\alpha(\chi)$ is the weak lensing radial kernel, and  $\eth$ is the spin-raising differential operator, acting on a spin-$s$ quantity as \citep{1966JMP.....7..863N}:
    \begin{equation}
      \eth\,_sf(\theta,\varphi)=-(\sin\theta)^s\left(\frac{\partial}{\partial\theta}+\frac{i}{\sin\theta}\frac{\partial}{\partial\varphi}\right)(\sin\theta)^{-s}\,_sf
    \end{equation}
    and turning it into a spin-$(s+1)$ quantity. The weak lensing kernel is\footnote{Note that this is only strictly valid in $\Lambda$CDM \citep{2019ARA&A..57..335F}.}
    \begin{align}\label{eq:wl_kernel}
      q_\alpha(\chi)\equiv\frac{3}{2}H_0^2\Omega_m\frac{\chi}{a(\chi)}\int_{z(\chi)}^\infty dz' p_\alpha(z')\frac{\chi(z')-\chi}{\chi(z')},
    \end{align}
    where $H_0\equiv H(z=0)$ is the Hubble expansion rate today, $\Omega_m$ is the current matter density parameter and, most importantly for our discussion, $p_\alpha(z)$ is the redshift distribution in bin $\alpha$,

    The angular power spectrum of the $E$-mode components of two maps $\alpha$ and $\beta$, $C_\ell^{\alpha\beta}$ can be related to the three-dimensional matter power spectrum $P(k,z)$ via:
    \begin{equation}\label{eq:limber}
      C_\ell^{\alpha\beta}=G_\ell^2\int\frac{d\chi}{\chi^2}\,q_\alpha(\chi)\,q_\beta(\chi)\,P\left(k=\frac{\ell+1/2}{\chi},z(\chi)\right),
    \end{equation}
    where we have assumed the Limber approximation \citep{1953ApJ...117..134L,2004PhRvD..69h3524A}, which is valid for the broad weak lensing kernels considered in this work. The scale-dependent lensing prefactor,
    \begin{equation}
      G_\ell\equiv\sqrt{\frac{(\ell+2)!}{(\ell-2)!}}\frac{1}{(\ell+1/2)^2},
    \end{equation}
    accounts for the difference between angular and three-dimensional derivatives in Eq. \ref{eq:deltagamma} (i.e. $\chi^2\eth^2\nabla^{-2}\not\equiv1$). This prefactor leads to sub-percent differences for $\ell>11$ and can therefore be neglected on small scales \citep{2017MNRAS.472.2126K}. In this work we will use the {\tt Halofit} fitting function of \cite{2003MNRAS.341.1311S,2012ApJ...761..152T} to describe the matter power spectrum.

    The intrinsic alignment (IA) of galaxies due to local interactions (gravitational or otherwise), is an important contaminant for cosmic shear data that must be taken into account \citep{astro-ph/0009499}. For simplicity, however, and since the focus of this work is the impact of the marginalisation over redshift distribution uncertainties, we will ignore the contribution from intrinsic alignments in this analysis.

  \subsection{Redshift distribution uncertainties}\label{ssec:methods.nzs}
    The sub-samples that make up the redshift bins used in the tomographic cosmic shear analysis of an imaging survey are selected based on the source photometry, either by simple cuts in the inferred photometric redshifts (photo-$z$), or by selecting directly in the magnitude-color space of the sample, bypassing photo-$z$ estimation altogether. Regardless of the method used to select the sub-samples, their true redshift distributions are inevitably subject to some level of uncertainty, due to the lack of precise redshift measurements.
    
    The $p(z)$ can however be calibrated through various methods, e.g.: weighted direct calibration with a sufficiently complete spectroscopic sample \citep{NIR_Nz1, NIR_Nz2}, clustering redshifts \citep{cross_corr_Nz1, cross_corr_Nz4, cross_corr_Nz2, cross_corr_Nz3}, and shear ratios \citep{DESY1_shear_ratios, DESY3_shear_ratios}. This typically leads to relatively tight priors on the $p(z)$, but the residual uncertainties in this prior must be propagated into the final parameter constraints.

    To characterise these uncertainties, we will make use of two different methods, which encompass the range of model complexity we may reasonably expect from current and future data.
    \begin{itemize}
      \item {\bf Method 1: $z$ shifts.} Most cosmic shear analyses to date \citep[among others]{HSC, KV450, K1K, DESY1, DESY3} have summarised the uncertainty in the calibrated $p_\alpha(z)$ into a single parameter $\Delta z^\alpha$ that shifts the mean of the redshift distribution. I.e. let $\hat{p}_\alpha(z)$ be the best-guess redshift distribution. The true redshift distribution is then
      \begin{equation}\label{eq:photo-z-model}
        p_\alpha(z)=\hat{p}_\alpha(z+\Delta z^\alpha).
      \end{equation}
      A prior on $\Delta z^\alpha$ can be derived using the calibration methods listed above. We will refer to this method as {\sl parametric}.

      This simple model turns out to be relatively well suited to describe the impact of $p(z)$ uncertainties in the case of cosmic shear data. Since weak lensing is a radially cumulative effect, the amplitude of the weak lensing kernel (Eq. \ref{eq:wl_kernel}) is mostly sensitive to the mean redshift of the sample, and thus much of the effect on cosmic shear observables is well described by this parameter \citep{2016PhRvD..94d2005B}.
      
      Other modes of $p(z)$ uncertainty, such as the distribution width, may be more relevant for galaxy clustering observables, or for the intrinsic alignment contribution to cosmic shear. Near-future cosmic shear samples may indeed require a more sophisticated description of the $p(z)$ uncertainty, and thus we turn to a more general method.

      \item {\bf Method 2: $p(z)$ bin heights.} Most $p(z)$ calibration methods (e.g. direct calibration or clustering redshifts) will produce a binned measurement of the $p(z)$ with deterministic redshift bin ranges, and uncertain bin heights. The most general method to propagate these uncertainties is therefore to treat each bin height $p_i\equiv p(z_i)$ as a free parameter in the model, with a prior given by the calibration uncertainties. The latter may be in the form of individual $1\sigma$ errors for each bin height, if the uncertainties are approximately uncorrelated, or a full covariance matrix covering all bin heights.

      The resulting parametrisation thus sidesteps any attempt at summarising the uncertainty into effective parameters, and thus we will refer to this method as {\sl non-parametric}. The method therefore fully propagates all calibration uncertainties into the final constraints with minimal approximations.
    \end{itemize}

    The key practical difference between both methods, in the context of error propagation, is the additional complexity they incur. The parametric approach (Method 1) introduces one free parameter per redshift bin. For $O(5)$ bins, this is already enough to significantly impact the performance of standard MCMC algorithms. In turn, the non-parametric approach (Method 2) introduces tens or hundreds of parameters per redshift bin, and one must resort to advanced sampling methods in order to fully explore the resulting model without assumptions.

  \subsection{Linearisation and analytical marginalisation}\label{ssec:methods.lin}
  
    Let $\pncal$ be the set of non-calibratable parameters of a model (in our case this is the set of cosmological and non-calibratable nuisance parameters) and $\pcal$ the set of calibratable parameter such that the total set of parameters is given by $\boldsymbol{\theta} = \pncal \cup \pcal $. Now consider the general case of a Gaussian posterior distribution of the form
    \begin{align}\nonumber
      -2\log P(\pncal,\pcal|{\bf d})=&({\bf d}-{\bf t})^T{\sf C}^{-1}({\bf d}-{\bf t})+(\pcal-\bar{\pcal})^T{\sf P}^{-1}(\pcal-\bar{\pcal})\\\label{eq:gauslike}
      &-2\log P(\pncal)+{\rm const.},
    \end{align}
    where ${\bf d}$ is the data. We assume a Gaussian calibration prior with mean $\bar{\pcal}$ and covariance ${\sf P}$, while $P(\pncal)$ is the prior on $\pncal$ (which is, as per our assumption, broad). ${\bf t}(\pncal,\pcal)$ is the theoretical prediction for the data ${\bf d}$ which implicitly depends on both calibratable and non-calibratable parameters. ${\sf C}$ is the covariance matrix of ${\bf d}$, which is parameter-independent.

    Assuming a tight prior on $\pcal$, we start by expanding the theory prediction around $\bar{\pcal}$
    \begin{equation}
      {\bf t}\simeq\bar{\bf t}+{\sf T}(\pcal-\bar{\pcal}),\hspace{6pt}{\rm where}\hspace{6pt}\bar{\bf t}\equiv{\bf t}(\pncal,\bar{\pcal}),\hspace{6pt}{\sf T}\equiv\left.\frac{d{\bf t}}{d\pcal}\right|_{\pcal=\bar{\pcal}}.
    \end{equation}
    Substituting this approximation in Eq. \ref{eq:gauslike}, the posterior becomes Gaussian in $\pcal$, and thus the calibratable parameters can be marginalised analytically. As shown in \citet{Boryana_Nz_2020}, the resulting marginalised posterior is
    \begin{align}\nonumber
      -2\log P(\pncal|{\bf d})\simeq&({\bf d}-\bar{\bf t})^T\tilde{\sf C}^{-1}({\bf d}-\bar{\bf t})-2\log P(\pncal)\\\label{eq:gauslike_marg}
      &+\log\left[\det\left({\sf T}^T{\sf C}^{-1}{\sf T}+{\sf P}^{-1}\right)\right]+{\rm const.},
    \end{align}
    where the modified covariance is
    \begin{equation}\label{eq:covmod}
      \tilde{\sf C}\equiv{\sf C}+{\sf T}{\sf P}{\sf T}^T.
    \end{equation}
    
    Note that, strictly speaking, both the modified covariance and the term in the second line of Eq. \ref{eq:gauslike_marg} depend on $\pncal$, which would in principle complicate the evaluation of the likelihood. In practice, this parameter dependence can be neglected such that the value of $\pncal$ at which these terms are evaluated can be fixed during exploration of the posterior. However, fixing $\pncal$ at values with a bad fit to the data will result in a mischaracterisation of the response of the theory vector to the nuisance parameters leading to inaccurate marginalised posteriors. Ideally, $\pncal$ is fixed to its maximum a posteriori (MAP) value. However, as shown in \citet{Boryana_Nz_2020} and in preliminary results, no appreciable differences are found in the marginalised posteriors for $\pncal$ within  2$\sigma$ of the MAP. Note that the size of the 2$\sigma$ region will depend on how constraining the data is.

    This result is intuitively simple to understand if we think of ${\sf T}$ as the response of the data to variations in the nuisance parameters. After marginalising over the calibratable parameters, the resulting distribution is a multi-variate Gaussian where the data covariance has been updated in Eq. \ref{eq:covmod} by increasing the uncertainty in the data modes that most prominently respond to variations in the nuisance parameters.
    \begin{figure*}
      \centering
      \includegraphics[width=0.95\textwidth]{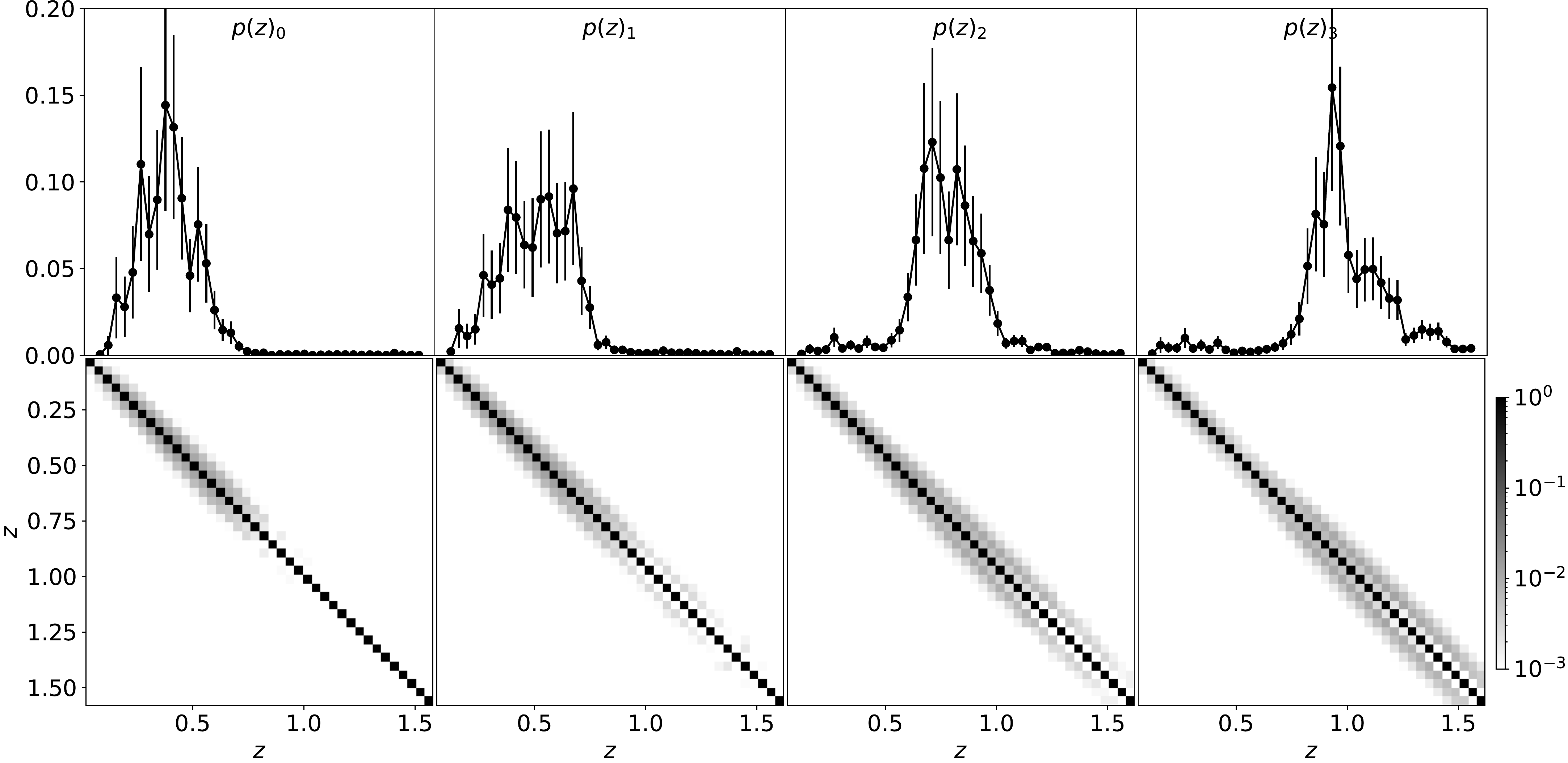}
      \caption{Top row: normalized galaxies's redshift distributions for each of the 4 redshift bins. Bottom row: correlation matrix obtained using the DIR algorithm for each of the 4 galaxies' redshift distributions. Note that for visualization purposes we display the absolute values of the each correlation matrix in logarithmic scale. In this plot we can see that the covariance matrices obtained through the DIR algorithm are mostly diagonal.} \label{fig:cov_pz}
    \end{figure*}

    In this work, $\pcal$ corresponds to the parameters describing the redshift distribution uncertainties, i.e. one shift parameter per redshift bin when using the parametric approach, or a set of $p(z)$ bin heights in the non-parametric scheme. The method described above, however, is fully general and has in the past been applied to marginalise over other types of nuisance parameters, including multiplicative shape measurement biases \citep{KV450}, as well as truly linear parameters such as shot-noise \citep{2021JCAP...10..030G} or systematic template amplitudes \citep{2020MNRAS.491.5464K}. The aim of this paper is thus to determine the applicability of this method to the case of redshift distribution uncertainties.

\section{Data}\label{sec:data}
  In order to evaluate the performance of the analytical marginalisation approach described in the previous section in the context of current and future surveys, we make use of data from the first-year cosmic shear analysis of the Dark Energy Survey (\desyo, \citet{2018ApJS..239...18A}). The aim of this is twofold: first, to demonstrate that the method can be successfully implemented in real data, with real-life complications (e.g. noisy $p(z)$s, numerical covariances, astrophysical and observational systematics) and, second, to demonstrate this validity for future Stage-IV datasets in the presence of $p(z)$ calibration uncertainties already achieved on current data. This section describes the \desyo data used, and the models used to generate simulated future Stage-IV data.

  \subsection{\desyo data and redshift distributions}\label{ssec:data.des}
    The Dark Energy Survey is a photometric, 5-year survey, that has observed 5000 deg$^2$ of the sky using five different filter bands (\textit{grizY}). The observations were made with the 4m Blanco Telescope, provided with the 570-Mpix Dark Energy Camera (DECam), from the Cerro Tololo Inter-American Observatory (CTIO), in Chile. In this paper we use cosmic shear data from the first data release \citep{2018ApJS..239...18A}, which covers 1786 deg$^2$ before masking. In particular, we use the public \mcal source catalog\footnote{\url{https://desdr-server.ncsa.illinois.edu/despublic/y1a1_files/}}, which is divided in four redshift bins covering the range $z\lesssim1.6$ \citep{2018MNRAS.478..592H}. 
    
    We use the calibrated redshift distributions of the \mcal sample provided by \citet{2023JCAP...01..025G}. The $p(z)$s were estimated via direct calibration (DIR \citet{NIR_Nz1}), using the COSMOS 30-band catalog \citep{2016ApJS..224...24L} as a calibrating sample. The uncertainties of the measured redshift distributions were estimated analytically, as described in \citet{2023JCAP...01..025G}, accounting for both shot noise and sample variance, and represent a realistic level of $p(z)$ uncertainty achieved by current existing datasets.  The redshift distributions were sampled on 40 bins of width $\delta z=0.04$ covering the range $0\leq z\leq1.6$. Fig. \ref{fig:cov_pz} shows, in the first row, the redshift distributions of the four \mcal samples and their statistical uncertainties. Note that we estimated the full covariance matrix of the $p(z)$ bin heights. The covariance is dominated by the diagonal, as can be seen in the bottom panels of Fig. \ref{fig:cov_pz}.

    We will also use the cosmic shear angular power spectra provided by \citet{2010.09717}. A full description of the methods used to estimate these power spectra, and their associated covariance matrix, from the \desyo data is provided by the authors.

  \subsection{Future Stage-IV data}\label{ssec:data.stage4}
    We generate a simulated data vector corresponding to a Stage-IV cosmic shear survey, such as the Legacy Survey of Space Time, at the Rubin Observatory \citep{2012arXiv1211.0310L}, or the Euclid survey \citep{2015arXiv150303757S}. Our aim is to effectively test the analytical marginalisation method in the low-noise regime, where the inferred posterior is likely more sensitive to residual $p(z)$ uncertainties, and the error budget may become dominated by these, rather than the statistical errors in the data themselves.

    For simplicity, we simulate the Stage-IV survey as having the same redshift distributions as the \desyo sample. This includes both the $p(z)$s themselves, and their calibration uncertainties. While it is possible that techniques for inferring redshifts from photometry, or the size and quality of calibrating spectroscopic samples, will improve substantially by the time Stage-IV data are available, we prefer to err on the side of caution and assume the same performance as currently achieved. For instance it is possible that redshift estimates will suffer commensurately with the increase in survey depth. The results presented here are therefore conservative, and their validity will only be reinforced if better $p(z)$ calibration samples are used in the future.

    We generate cosmic shear power spectra using \ccl \citep{ccl} for the best-fit {\it Planck} 2018 cosmological parameters \citep{2020A&A...641A...6P}: $\Omega_b h^2 = 0.02237$, $\Omega_c h^2 = 0.12$, $h = 0.6736$, $10^9 A_s = 2.0830$, $n_s = 0.9649$, $w_0 = -1$, $w_a = 0$. We use the same sampling in $\ell$ used for the \desyo power spectra, and use only scales in the range $\ell\in[30,2000]$.

    We compute the covariance matrix of these power spectra analytically, including a disconnected ``Gaussian'' component, and a connected super-sample covariance contribution (SSC).
    \begin{equation}
      {\rm Cov}\left(C^{\alpha}_\ell, C^{\rho\sigma}_{\ell'}\right) = {\rm Cov}_G\left(C^{\alpha\beta}_\ell, C^{\rho\sigma}_{\ell'}\right) + {\rm Cov}_{\rm SSC}\left(C^{\alpha\beta}_\ell, C^{\sigma\rho}_{\ell'}\right).
    \end{equation}
    We estimate the Gaussian covariance using a simple mode-counting approximation  \citep{2004MNRAS.349..603E} as
    \begin{equation}
      {\rm Cov}_G\left(C^{\alpha\beta}_\ell, C^{\rho\sigma}_{\ell'}\right) = \delta_{\ell\ell'}\frac{C^{\alpha\rho}_\ell C_\ell^{\beta\sigma}+C_\ell^{\alpha\sigma} C_\ell^{\beta\rho}}{(2\ell+1)\,\Delta\ell\,f_{\rm sky}},
    \end{equation}
    where $f_{\rm sky}$ is the fraction of the sky covered by the experiment. We assume $f_{\rm sky}=0.4$, as in the case of LSST \citep{2012arXiv1211.0310L}. The angular power spectra above contain the contribution from shape noise in the auto-correlation, of the form
    \begin{equation}
      N_\ell^{\alpha\alpha}=\frac{\sigma_\gamma^2}{\bar{n}_\alpha}.
    \end{equation}
    Here $\sigma_\gamma=0.28$ is the per-component ellipticity dispersion in each source, and $\bar{n}_\alpha$ is the angular number density of sources in the $\alpha$-th redshift bin. We assume $n_\alpha=4\,{\rm arcmin}^{-2}$ in each redshift bin.
    
    We compute the super-sample covariance contribution following: 
    \begin{align}
      \mathrm{Cov}_{\mathrm{SSC}}(C^{\alpha\beta}_{\ell}, C^{\rho\sigma}_{\ell'}) = &\int \mathrm{d}\chi \;\frac{q^\alpha(\chi)q^\beta(\chi)q^\rho(\chi)q^\sigma(\chi)}{\chi^{4}} \times \\
      &\frac{\partial P(\ell/\chi, z)}{\partial \delta_{\rm LS}}\frac{\partial P(\ell'/\chi, z)}{\partial \delta_{\rm LS}}\sigma^{2}_{\rm LS}(z) ,
    \end{align}
    as in \citet{2010.09717}. $\partial P(k, z)/\partial \delta_{\rm LS}$ is the response of the matter power spectrum to a large-scale density fluctuation $\delta_{\rm LS}$, and the quantity $\sigma_b^2(z)$ is the variance of the long wavelength mode over the survey footprint. We estimate the latter as in \citet{2017MNRAS.470.2100K}, modelling the footprint simply as a circular cap of area $4\pi f_{\rm sky}$. We estimate the response function using perturbation theory and the halo model, as described in \citet{2017MNRAS.470.2100K}, and as implemented in \ccl.

  \section{Likelihood}\label{sec:likelihood}
    \begin{table}
      \centering
      \def\arraystretch{1.2}
      \begin{tabular}{|cc|cc|}
        \hline
        \multicolumn{4}{|c|}{\textbf{Parameter priors}} \\
        \hline
        Parameter &  Prior & Parameter &  Prior\\  
        \hline 
        \multicolumn{2}{|c|}{\textbf{Cosmology}}  &        \multicolumn{2}{|c|}{\textbf{Redshift calibration}} \\
        $\Om$  &  $U (0.1, 0.9)$  &             $\Delta z_1 $  & $\cN 0.0,0.016)$ \\ 
        $\Ob$  &  $U (0.03, 0.07)$  &            $\Delta z_2 $  & $\cN (0.0,0.017)$  \\
        $h$   &  $U (0.55, 0.91)$  &            $\Delta z_3 $  & $\cN (0.0, 0.013)$ \\
        $\ns$ & $U (0.87, 1.07)$  &             $\Delta z_4 $  & $\cN (0.0, 0.015)$ \\
        $\sig$ &  $U(0.6, 0.9)$   &             $p_{{\rm i}} $  & $\cN (\bar{p}_i, {\sf C})$  \\ \multicolumn{2}{|c|}{\textbf{Shear multiplicative bias }}& &\\ 
        $m^i $ & 0.012  &  & \\
        \hline
      \end{tabular}
      \caption{Prior distributions for the parameters considered in this work. Note that the redshift calibration section contains the priors for both the $\Delta z$ and $p_{\alpha}(z)$ models which are not sampled simultaneously.}\label{tab:priors}
    \end{table}
    We extract cosmological parameter constraints using a Gaussian likelihood as described in Section \ref{ssec:methods.lin}. In order to validate the analytical marginalisation approach, we will either use the full posterior distribution in Eq. \ref{eq:gauslike}, or the analytically marginalised version in Eq. \ref{eq:gauslike_marg}\footnote{Recall that we treat the term in the second line of Eq. \ref{eq:gauslike_marg} as a constant.}. In the first case, $\pcal$ includes all nuisance parameters describing the redshift distribution uncertainties, and in both cases $\pncal$ includes all other model parameters. Specifically, $\pncal$ contains the five $\Lambda$CDM cosmological parameters $(\Omega_{\rm m},\Omega_{\rm b},\sigma_8,n_s,h)$.

    When marginalising over redshift distribution uncertainties, $\pcal$ will contain either one redshift shift parameter $\Delta z_\alpha$ for each redshift bin, when employing the parametric description of $p(z)$ uncertainties (Method 1), or a set of bin heights for each redshift bin determining $p_\alpha(z)$, when using the non-parametric approach (Method 2). The first case will introduce $4$ new parameters to the model, while the latter will introduce $4\times40=160$ new amplitude parameters, as described in Section \ref{ssec:data.des}.

    Table \ref{tab:priors} shows the parameter priors used in this work. All cosmological parameters take uniform, largely uninformative priors. For simplicity, the multiplicative bias parameters were fixed at the center of the Gaussian priors from the official analysis of \desyo \citep{DESY1}. When using Method 1 to numerically marginalise over the $p(z)$ uncertainties, we used Gaussian priors on each of the shift parameters $\Delta z_\alpha$, following those used by \desyo \citep{DESY1}. When using Method 2 (marginalisation over $p(z)$ bin amplitudes), we assume a multi-variate Gaussian prior, with the $p(z)$ covariance described in Sect. \ref{ssec:data.des} and shown in Fig. \ref{fig:cov_pz}. 

    For both $p(z)$ uncertainty models, when using analytical marginalisation, we use Eq. \ref{eq:gauslike_marg} and modify the covariance as in Eq. \ref{eq:covmod}, with ${\sf P}$ given by the priors described above. When using numerical marginalisation, we simply explore the posterior distribution of the full model, including all the $p(z)$, $p_i$, parameters. In the case of Method 2, this involves sampling a distribution with 165 parameters, of which the bulk (160 parameters) describe the $p(z)$ uncertainty. This is not feasible for standard Metropolis-Hastings MCMC methods \cite{Metropolis, Hastings} due to the curse of dimensionality, and therefore we resort to a Hamiltonian Monte Carlo (HMC) approach.
    
    HMC \citep{HMC, Betancourt17} uses notions of Hamiltonian dynamics to draw trajectories on the parameter space along which the sampler moves. This results in a much greater acceptance rate, and allows HMC to beat the dimensionality curse. HMC can thus efficiently explore parameter spaces with large numbers of dimensions in far less time than Metropolis-Hastings or nested sampling techniques \citep{NS_Handley}. The main difficulty of using HMC is the need to calculate gradients of the log-posterior to calculate the Hamiltonian equations of motion. The additional computational cost of obtaining these derivatives numerically (e.g. via adaptive finite differences) may outweigh the gains caused by the higher acceptance rates of HMC. To overcome this problem we make use of automatic differentiation (AD). To take advantage of AD, we have developed a cosmological theoretical prediction code natively written in the {\tt Julia} programming language \citep{LimberJack}. \texttt{Julia} is a just-in-time (JIT) compiled language with C-like performance and seamless AD integration, which can thus be used to efficiently sample complex cosmological posteriors using HMC. To sample the posterior distribution we use the No-U-Turns Sampler (NUTS \citet{NUTS}) implementation of HMC within the {\tt Turing.jl} package \citep{ge2018t}.

  \section{Results}\label{sec:results}
    \subsection{Linearising \texorpdfstring{$\Delta z$}{Lg}}\label{ssec:results.dz}
      Let us begin the discussion of our results by considering the simplest of the two models of the photometric uncertainties studied in this work, the $\Delta z$ model (called Method 1 above). As discussed in Section \ref{sec:likelihood}, this model introduces 4 new shift parameters $\Delta z$ (one per redshift bin) in addition to the 5 \lcdm parameters. All other nuisance parameters are kept fixed. For the \desyo and LSST-like datasets, we will compare the result of analytically marginalizing over the $\Delta z$ parameters against performing the full numerical marginalisation on the corresponding cosmological constraints. In order to quantify the contribution of redshift uncertainties to the total error budget, we will also present results for the case when the $\Delta z$ parameters are fixed (i.e. assuming perfect knowledge of the redshift distributions).

\begin{figure}
  \centering
  \includegraphics[width=0.95\linewidth]{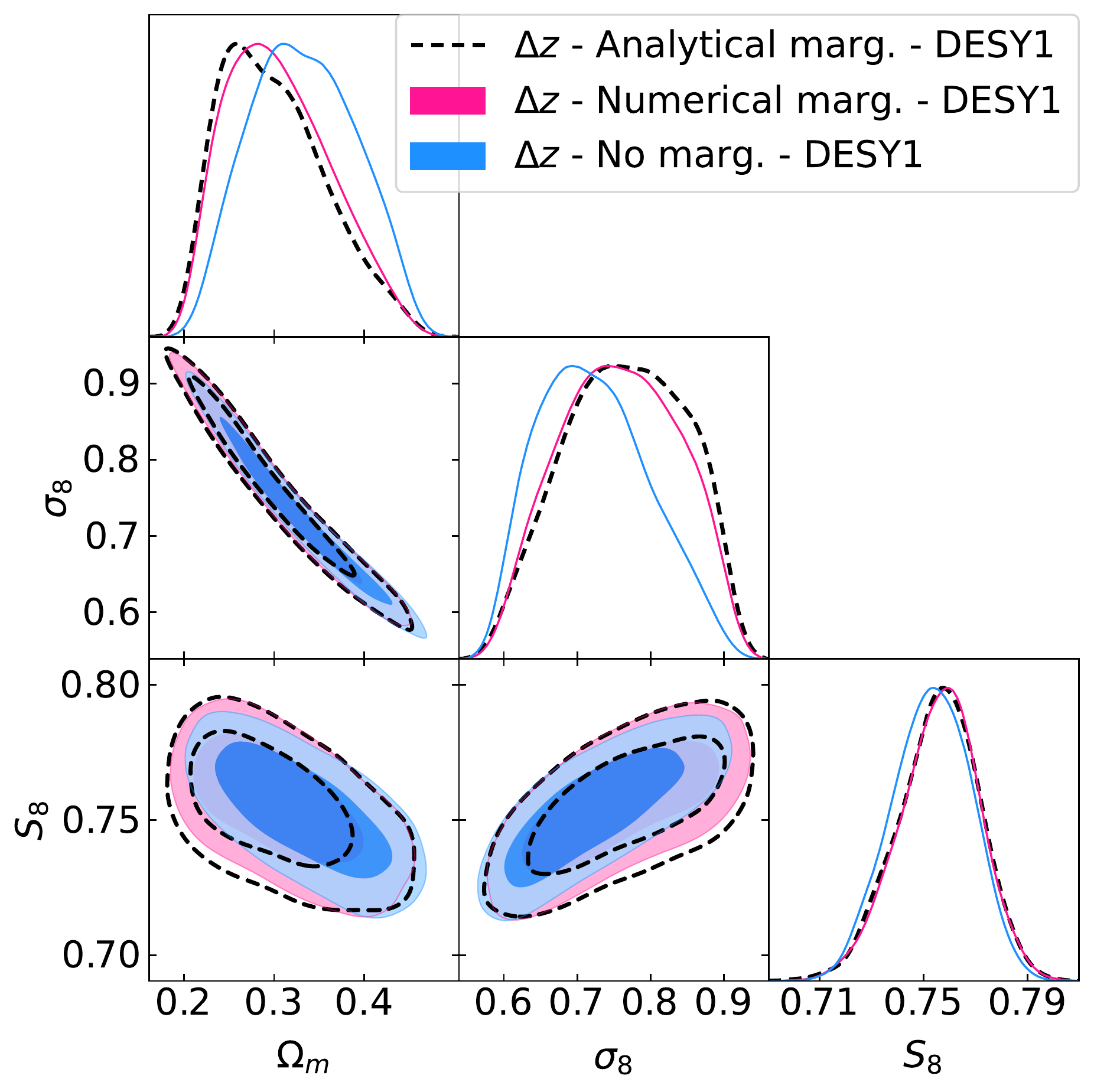}
  \caption{Marginalised posterior distributions for the combination of parameters $\Omega_{\rm{m}}$, $\sigma_{\rm{8}}$ and $S_{\rm{8}}$ obtained when considering the $\Delta z$ model for photometric uncertainties for \desyo data. The blue contours correspond to the case where the $\Delta z$ parameter are fixed. The magenta contours are obtained when numerically marginalizing over the $\Delta z$ parameters. Finally, the black dashed contours are obtained when analytically marginalizing over the $\Delta z$ parameters. We can observe that the analytical and numerical marginalisation return nearly identical posteriors.}
  \label{fig:dz_DESY1}
\end{figure}

  Our results for \desyo data are shown in  Fig. \ref{fig:dz_DESY1}, with the errors on all parameters listed in Table \ref{tab:dz_cosmo_constraints}. On the one hand, we find that marginalizing analytically or numerically over the $\Delta z$ parameters leads to the same marginalised posterior for the cosmological parameters. On the other hand, fixing the $\Delta z$ parameters returns a posterior distribution that is only mildly narrower than the marginal distribution. For the \desyo data, the impact of redshift uncertainties in the final cosmological errors is relatively small (although not negligible). Thus, if we truly wish to study the effect of marginalizing analytically as opposed to numerically over the $\Delta z$ parameters we will have to consider futuristic LSST-like data, where the impact of these uncertainties will likely be higher. 

\begin{figure}
  \centering
  \includegraphics[width=0.95\linewidth]{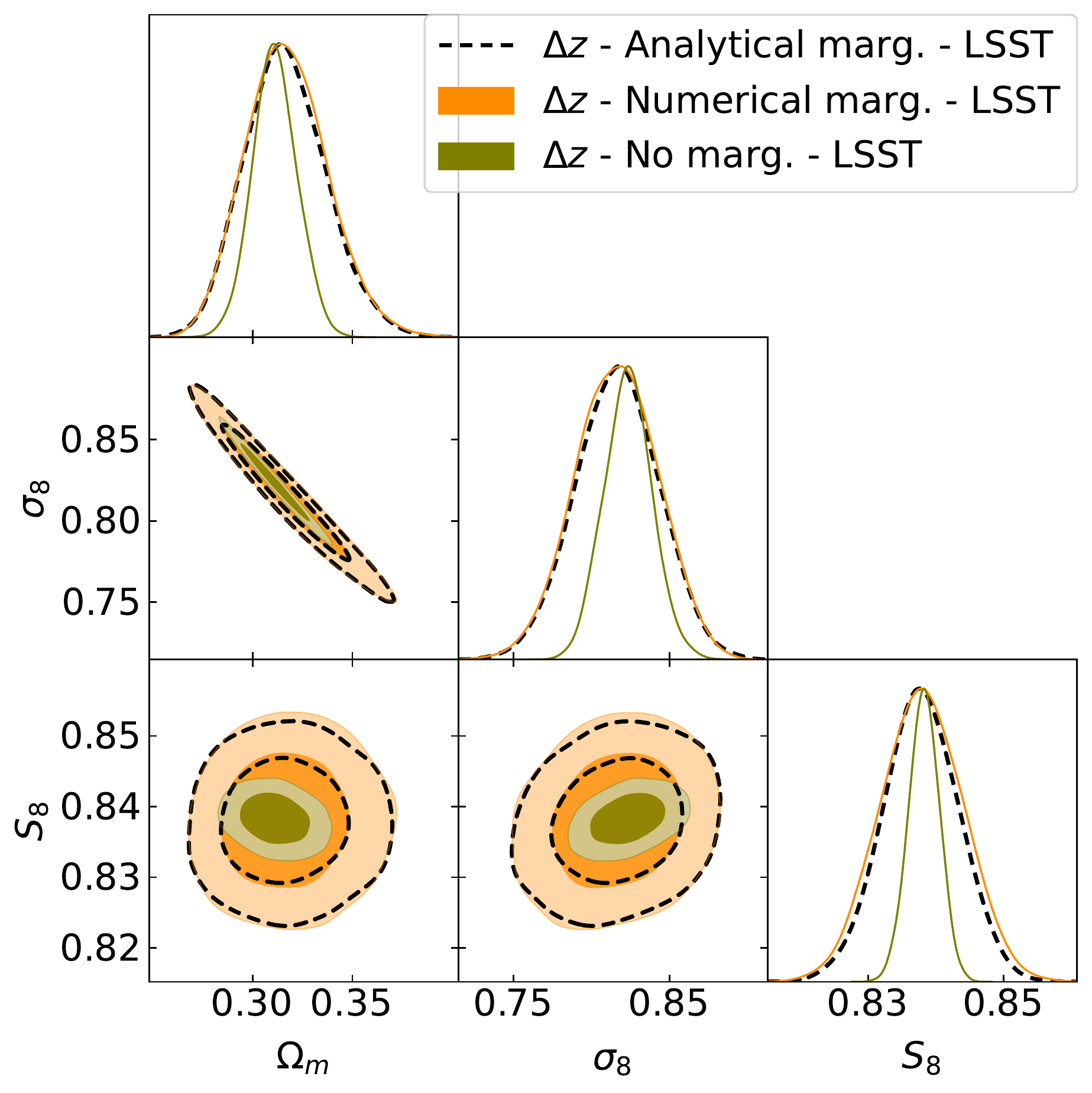}
  \caption{Marginalised posterior distributions for the combination of parameters $\Omega_{\rm{m}}$, $\sigma_{\rm{8}}$ and $S_{\rm{8}}$ obtained when considering the $\Delta z$ model for photometric uncertainties for futuristic LSST-like data. The green contours correspond to the case where the $\Delta z$ parameter are fixed. The orange contours are obtained when numerically marginalizing over the $\Delta z$ parameters. Finally, the black dashed contours are obtained when analytically marginalizing over the $\Delta z$ parameters. We can observe that the analytical and numerical marginalisations return nearly identical posteriors.}
  \label{fig:dz_LSST}
\end{figure}

  We show results for futuristic LSST-like data on Fig. \ref{fig:dz_LSST}, with the parameter constraints listed in Table \ref{tab:dz_cosmo_constraints}. First of all, in the case LSST-like data we observe that not marginalising over the $\Delta z$ parameters in the model results in significantly narrower posteriors, with the final uncertainties shrinking by a factor $\sim2$. The impact of redshift distribution uncertainties in this case is thus much more relevant, and the accuracy of the analytical marginalisation scheme becomes paramount. However, comparing the contours obtained by numerical and analytical marginalisation, we observe that both methods return largely equivalent posterior distributions, with the final uncertainties changing by much less than $10\%$. This holds even in the case the $\Delta z$ prior worsen by a factor 4 as seen in Figure~\ref{fig:4x}, in Appendix~\ref{app:approx}. Therefore, linearizing the likelihood around the $\Delta z$ parameters will be a good enough approximation for LSST-data, at least for relatively simple parametrisations of the $p(z)$ uncertainty, which will allow us to reduce the dimensionality of the model and make parameter inference more efficient. 

  It is worth emphasizing that the results in this section are not meant to be interpreted as forecasts on the constraining power of LSST on cosmological parameters, but only on our ability to analytically marginalize over photometric uncertainties in inferring the underlying cosmology. The recovered constraints depend strongly on assumptions such as the redshift calibration that LSST will be able to achieve for the different samples involved. As such, the results presented here are only a conservative estimate of the effect of analytic marginalisation on cosmological constraints.

\begin{table} 
\centering
\begin{tabular}{|c|c|c|c|c}

\multicolumn{2}{c}{$\Delta z$ model} & Fixed & Numerical & Analytical  \\ 
\hline
\multirow{2}{*}{$\Omega_{\rm{m}}$} & \desyo &  0.333  $\pm$ 0.055 & 0.3 $\pm$ 0.056 & 0.306 $\pm$ 0.055 \\ \cline{2-5}
                                   & LSST  & 0.311  $\pm$ 0.011 & 0.317 $\pm$ 0.02 & 0.317 $\pm$ 0.02 \\
\hline
\multirow{2}{*}{$\sigma_{\rm{8}}$} & \desyo & 0.724 $\pm$ 0.072 & 0.765 $\pm$ 0.077 & 0.758 $\pm$ 0.076 \\ \cline{2-5}
                                   & LSST  & 0.82 $\pm$ 0.015 & 0.821 $\pm$ 0.027 & 0.823 $\pm$ 0.027 \\
\hline
\multirow{2}{*}{$S_{\rm{8}}$} & \desyo & 0.753 $\pm$ 0.015 & 0.756 $\pm$ 0.015 & 0.756 $\pm$ 0.015 \\ \cline{2-5}
                              & LSST  & 0.833 $\pm$ 0.002 & 0.833 $\pm$ 0.005 & 0.833 $\pm$ 0.006  \\
\hline
\end{tabular}
\caption{Numerical values for the mean and 1$\sigma$ confidence intervals for the 1D marginalised posterior distributions of the cosmological parameters $\Omega_{\rm{m}}$, $\sigma_{\rm{8}}$ and $S_{\rm{8}}$ obtained when considering the first method ($z$ shifts) to characterise the photometric redshift uncertainties. The first column shows the values obtained when the $\Delta z$ parameters were kept fixed, the second column when they were marginalised numerically and the third column when they were marginalised analytically. In each row we display the constraints obtained when using \desyo or LSST-like data to constrain the models. }
\label{tab:dz_cosmo_constraints}
\end{table}

\subsection{Linearising \texorpdfstring{$p_{\alpha}(z)$}{Lg}}\label{ssec:results.Nz}
  In the previous section we have shown that, even for futuristic LSST-like data, it is possible to marginalize over redshift uncertainties analytically, assuming a relatively simple parametrisation of these uncertainties. We now turn to more complex models to characterise these uncertainties.

  In order to do so we consider the previously discussed $p_{\alpha}(z)$ model (called Method 2 above), which turns the height of each bin in the redshift distribution histograms into a free parameter. This results in 40 new free parameters per redshift bin with a total of 160 parameters for the data considered in this work.

\begin{figure}
  \centering
  \includegraphics[width=0.95\linewidth]{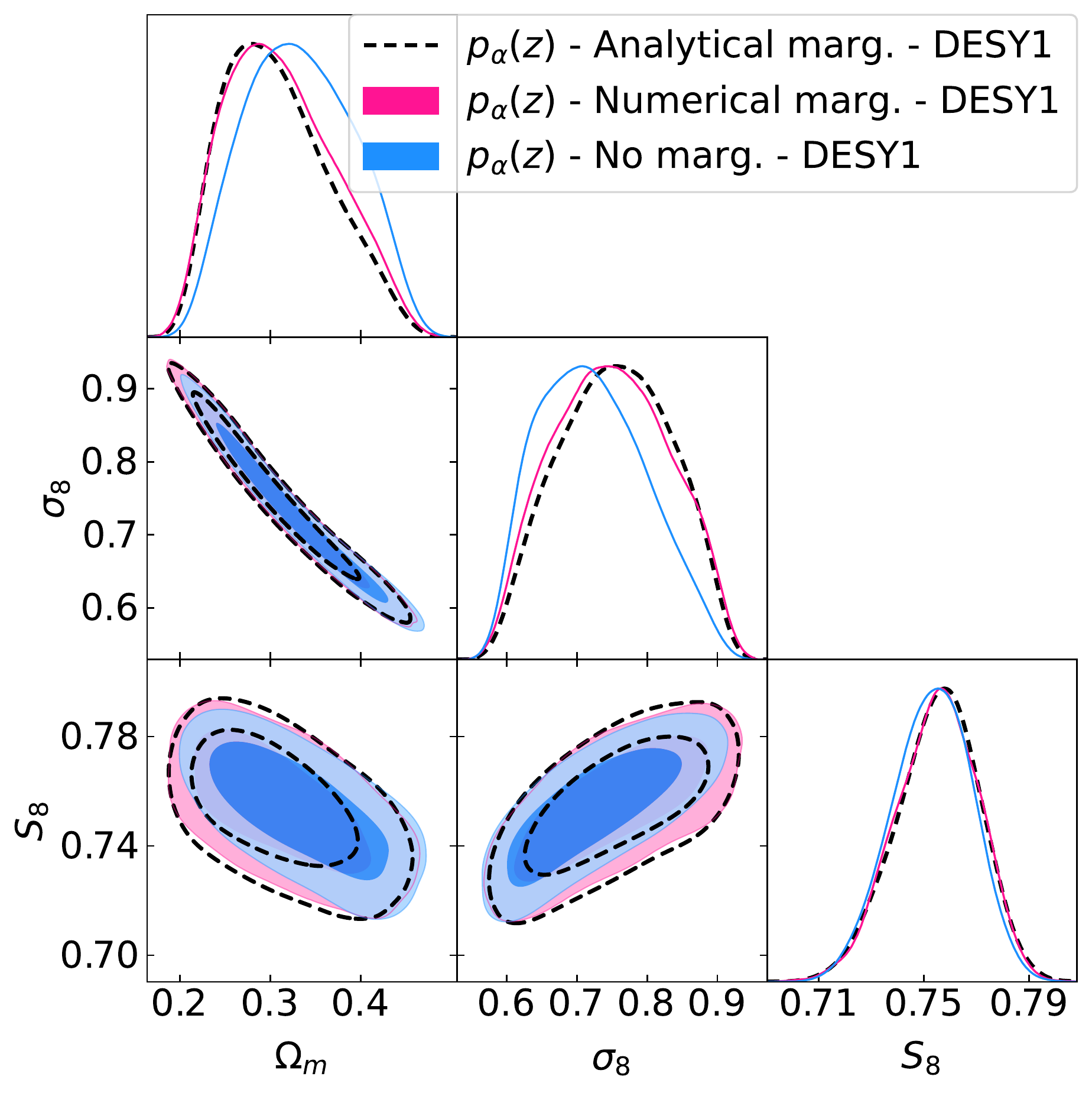}
  \caption{Marginalised posterior distributions for the combination of parameters $\Omega_{\rm{m}}$, $\sigma_{\rm{8}}$ and $S_{\rm{8}}$ obtained when considering the $p_{\alpha}(z)$ model for photometric uncertainties for \desyo data. The blue contours correspond to the case where the $p_{\alpha}(z)$ parameter are fixed. The magenta contours are obtained when numerically marginalising over the $p_{\alpha}(z)$ parameters. Finally, the black dashed contours are obtained when analytically marginalizing over the $p_{\alpha}(z)$ parameters. We can observe that the analytical and numerical marginalisation return nearly identical posteriors.}
  \label{fig:Nz_DESY1}
\end{figure}

  We start by revisiting the \desyo data analysis, presenting our results in Fig. \ref{fig:Nz_DESY1}. As we observed in the previous section, we find that even for the far more general $p_{\alpha}(z)$ model there is no significant difference between numerically marginalizing over the $p_{\alpha}(z)$, or doing so through our approximate analytical approach. Furthermore, as before, fixing the shape of the redshift distribution leads to only mildly tighter constraints. On the one hand, this means that the result found for the $\Delta z$ model is not reliant on the simplicity of the model, but instead inherent to the sensitivity of \desyo data. On the other hand, this also means that we must turn once again to futuristic LSST-like data to study the impact of a more general parametrisation of photometric uncertainties. 

\begin{figure}
  \centering
  \includegraphics[width=0.95\linewidth]{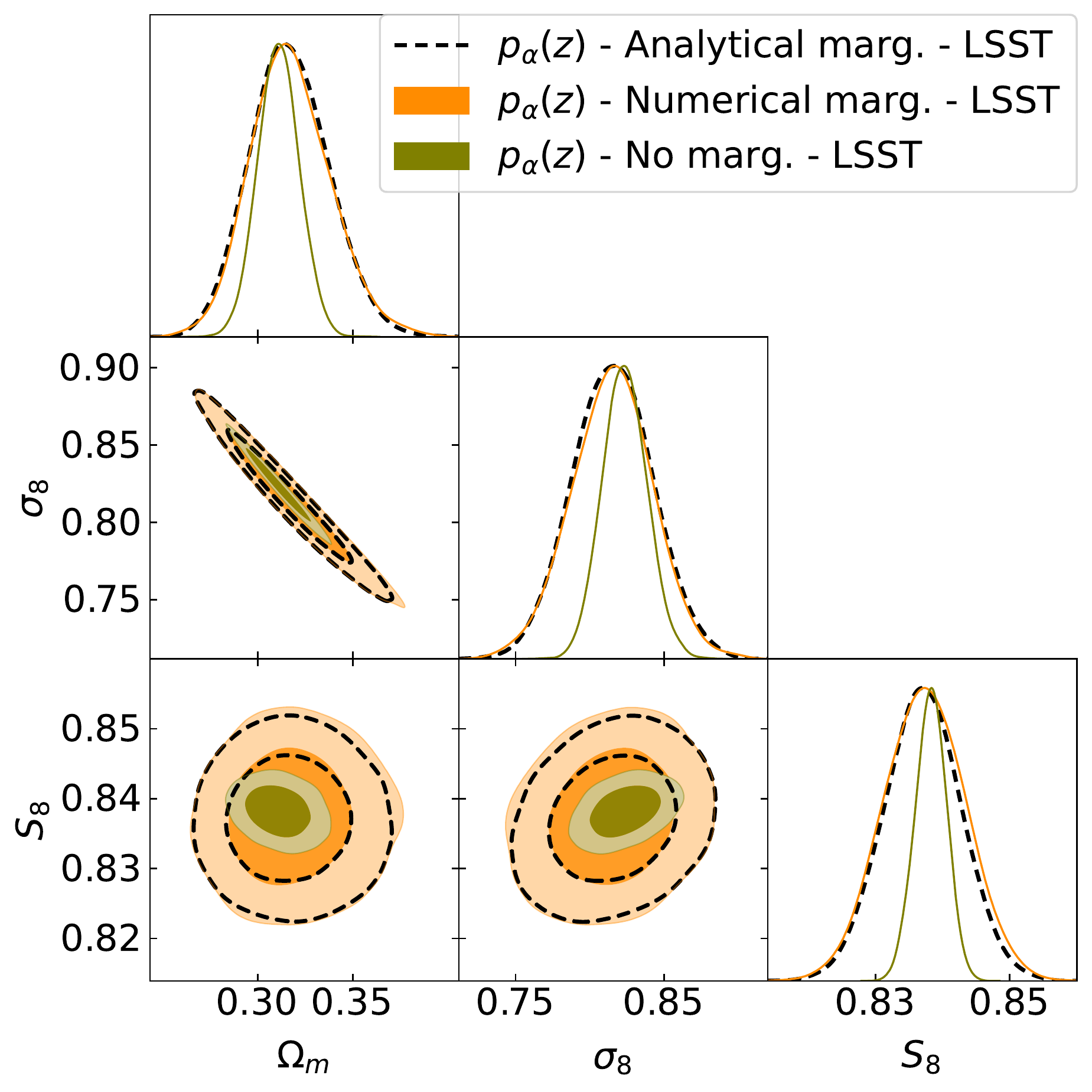}
  \caption{Marginalised posterior distributions for the combination of parameters $\Omega_{\rm{m}}$, $\sigma_{\rm{8}}$ and $S_{\rm{8}}$ obtained when considering the $p_{\alpha}(z)$ model for photometric uncertainties for LSST-like futuristic data. The green contours correspond to the case where the $p_{\alpha}(z)$ parameter are fixed. The orange contours were obtained when numerically marginalizing over the $p_{\alpha}(z)$ parameters. Finally, the black dashed contours were obtained when analytically marginalizing over the $p_{\alpha}(z)$ parameters. We can observe that the analytical and numerical marginalization return nearly identical posteriors. }
  \label{fig:Nz_LSST}
\end{figure}

The results for futuristic LSST-like data are shown in Fig. \ref{fig:Nz_LSST}. As in the case of the $\Delta z$ parametrisation, we find that, in the case LSST-like data, not including the $p_{\alpha}(z)$ parameters in the model results in significantly narrower posteriors. By looking at the corresponding numerical values in Tab. \ref{tab:Nz_cosmo_constraints}, we see that the $S_8$ constraints become twice as tight when the $p_{\alpha}(z)$ parameters are fixed. Most importantly, we find that marginalizing over the $p_{\alpha}(z)$ parameters analytically or numerically yields almost indistinguishable posteriors. Thus, the results found in Sect. \ref{ssec:results.dz} for the simple $\Delta z$ parametrisation, in fact hold for significantly more general models of the uncertainty in the galaxy redshift distributions.

\begin{table} 
\centering
\begin{tabular}{|c|c|c|c|c}
 \multicolumn{2}{c}{$p_{\alpha}(z)$ model} & Fixed & Numerical & Analytical  \\ 
 \hline
\multirow{2}{*}{$\Omega_{\rm{m}}$} & \desyo &  0.333  $\pm$ 0.056 & 0.308 $\pm$ 0.055 & 0.312 $\pm$ 0.057 \\ \cline{2-5}
                                   & LSST  & 0.311  $\pm$ 0.011 & 0.317 $\pm$ 0.02 & 0.317 $\pm$ 0.021 \\
  \hline
\multirow{2}{*}{$\sigma_{\rm{8}}$} & \desyo & 0.723 $\pm$ 0.073 & 0.755 $\pm$ 0.075 & 0.75 $\pm$ 0.077 \\ \cline{2-5}
                                   & LSST  & 0.824 $\pm$ 0.015 & 0.816 $\pm$ 0.026 & 0.815 $\pm$ 0.027 \\
  \hline
\multirow{2}{*}{$S_{\rm{8}}$} & \desyo & 0.753 $\pm$ 0.015 & 0.755 $\pm$ 0.015 & 0.755 $\pm$ 0.015 \\ \cline{2-5}
                              & LSST  & 0.838 $\pm$ 0.002 & 0.837 $\pm$ 0.006 & 0.837 $\pm$ 0.006  \\
  \hline
\end{tabular}
\caption{Numerical values for the mean and 1$\sigma$ confidence intervals for the 1D marginalised posterior distributions of the cosmological parameters $\Omega_{\rm{m}}$, $\sigma_{\rm{8}}$ and $S_{\rm{8}}$ obtained when considering the second method ($p(z)$ bin heights) to characterise the photometric redshift uncertainties. The first column shows the values obtained when the $p_{\alpha}(z)$ parameters are kept fixed, the second column when they are marginalised numerically, and the third column when they are marginalised analytically. In each row we display the constraints obtained when using \desyo or LSST-like data to constrain the models.}
\label{tab:Nz_cosmo_constraints}
\end{table}

\begin{figure}
  \centering
  \includegraphics[width=0.5\textwidth]{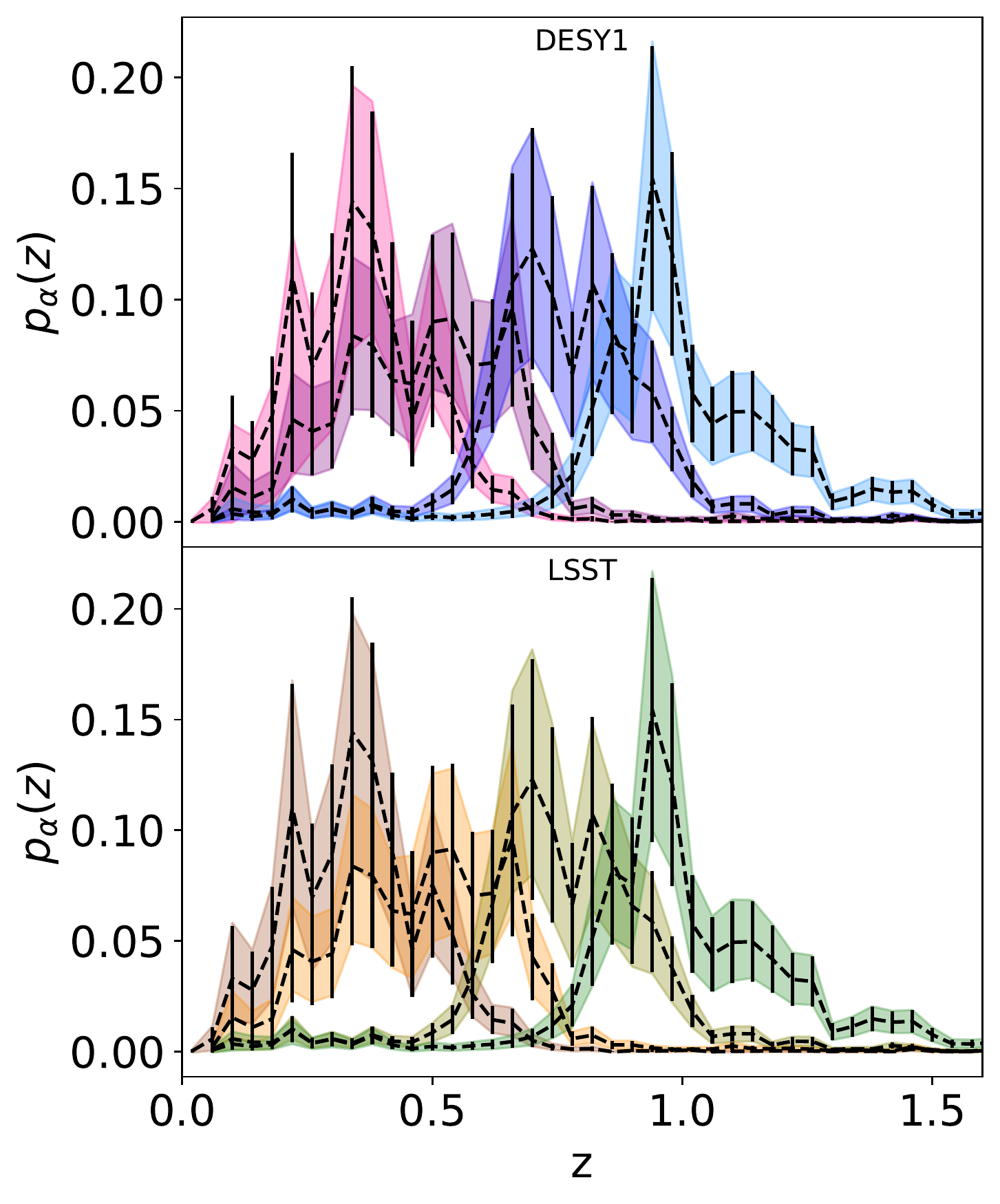}
  \caption{Posterior distributions for the $p_{\alpha}(z)$ parameters when considering \desyo data (top row) and futuristic LSST-like data (bottom row). The black dashed line shows the mean of the Gaussian prior of the $p_{\alpha}(z)$ parameters. The error bars show their corresponding error.}
  \label{fig:Nzs}
\end{figure}

Finally, in Fig. \ref{fig:Nzs} we present the constraints obtained for the 160 $p_{\alpha}(z)$ parameters for both the \desyo (top panel) and LSST-like data (bottom panel) in color bands. We observe that the posterior distributions are largely dominated by the prior (shown in dashed black line with error bars) and, thus, the redshift distribution is not significantly self-calibrated by the data in either case. 

Before moving to the next Section, it is worth stressing that constraining such a large parameter space has only been possible thanks to the auto-differentiable nature of the code used to obtain theoretical predictions, allowing us to use gradient-based samplers, much more efficient that standard samplers. The development of such auto-differentiable codes will therefore become imperative in the near future given the increasing complexity of models used in cosmological analyses.

\subsection{\texorpdfstring{$\Delta z$ vs $p_{\alpha}(z)$}{Lg}}\label{ssec:results.vs}

In the previous sections we have focused in the impact of how we marginalize over the different parametrisations of photometric redshift uncertainties. In this section we will focus instead on what we marginalize over, i.e. the impact of the choice of parametrisation. The question is then: Can a one-parameter-per-bin model ($\Delta z$ model) capture all the meaningful modifications to photometric redshift distributions?

\begin{figure*}
  \centering
  \includegraphics[width=0.95\textwidth]{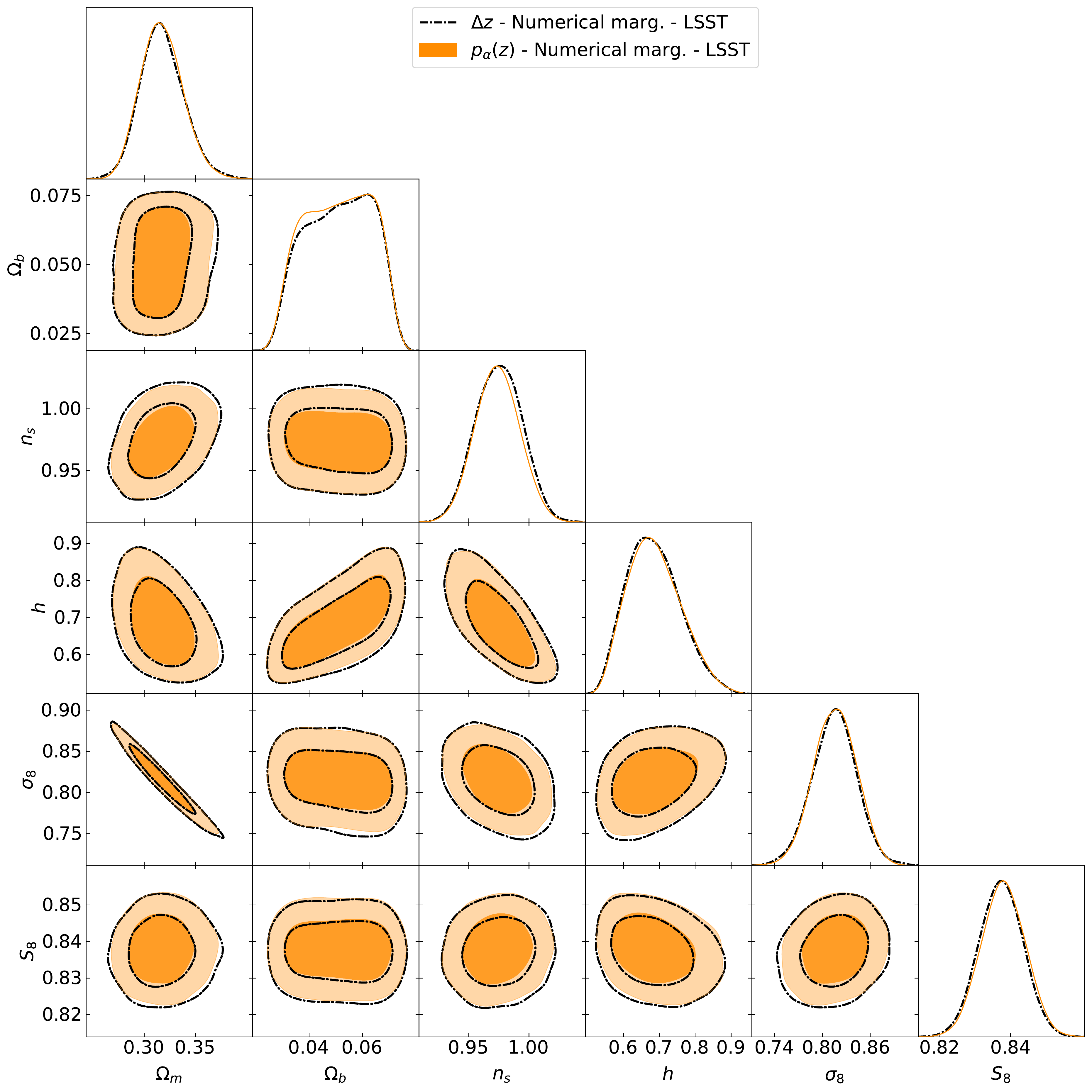}
  \caption{Comparison between the obtained marginalised posterior distributions of the cosmological parameters when numerically marginalizing over the $\Delta z$ (black dash-dotted) and $p_{\alpha}(z)$ (orange) photometric uncertainties models  when applied to LSST-like futuristic data. We can observe that both prametrizations of the photometric redshift uncertainties return identical posteriors for the cosmological parameters.}
  \label{fig:comp_dz_nz}
\end{figure*}

In order to answer this question, we constrain the cosmological parameters for the $\Delta z$ and $p_{\alpha}(z)$  models in the case with futuristic LSST-like data. In both cases, we marginalize numerically over their respective nuisance parameters. As shown in Fig. \ref{fig:comp_dz_nz} and Tables \ref{tab:dz_cosmo_constraints} and \ref{tab:Nz_cosmo_constraints}, both methods recover the same posterior distributions with small differences. Thus, it is in principle possible that even Stage-IV surveys will be able to use relatively simple models to describe the redshift distribution of cosmic shear samples\footnote{Note, however, this is likely not the case for photometric galaxy clustering studies where other properties of the redshift distribution (e.g. its width) have a stronger impact on the theoretical prediction \citep{2020JCAP...03..044N}.}. 

\section{Conclusions}\label{sec:conc}
One of the most significant obstacles to overcome in photometric weak lensing surveys is the accurate modeling of redshift distributions, $p(z)$. Not only are our measurements prone to error, which can bias the inferred cosmological parameters, but accounting for these uncertainties is also  a major inhibitor of efficient parameter inference. In this paper, we investigate the impact of analytically marginalizing over the uncertainties in the redshift distribution of galaxies in weak lensing surveys, as initially proposed in \citet{Boryana_Nz_2020}. In particular, we thoroughly quantify the validity of this approach for a current weak lensing survey, DES, as well as for a futuristic LSST-like survey, testing whether a fast analytic method proposed in this work is capable of reproducing the posterior distributions and constraints one arrives at when adopting the traditional method of diligently varying tens or hundreds of nuisance parameters.

Our results show that, for present surveys, marginalizing over the uncertainty in the redshift distribution of galaxies has only a mild impact on the constraints on cosmological parameters, although one that our analytical approximation is able to reproduce accurately. This is true for the two parametrisations of the uncertainties considered in this work, in terms of mean redshift shifts or redshift distribution histogram heights. However, the impact of redshift distribution uncertainties changes dramatically for future LSST-like surveys. In this case, redshift uncertainties commensurate with current calibration samples lead to an degradation in the final constraints on cosmological parameters of up to a factor $\sim2$. Capturing this effect for an arbitrarily complex parametrisation of the redshift distribution uncertainties is an a priori difficult task without resorting to a full exploration of the parameter space. Nevertheless, we find that the analytical approximate scheme explored here is still able to recover the marginalised constraints on cosmological parameters to high fidelity, even after marginalising over more than 100 nuisance parameters. This means that, while future surveys will certainly have to account for these uncertainties, they will be able to do so using fast marginalisation methods without increasing the dimensionality of their astrophysical and cosmological models.

Our results have also shown that simple parametrisations of the redshift distribution for cosmic shear samples, in terms of shifts in the mean redshift, are, surprisingly, able to reproduce the impact of the full uncertainty on $p(z)$ on the final constraints to high precision. Although this result will likely not hold for other probes (e.g. tomographic galaxy clustering), it should certainly simplify the analysis of future cosmic shear data.

It is worth emphasizing that our work has focused exclusively on the case of cosmic shear data, and that our conclusions only apply in this context. The validity of the analytical approximation employed here for general tomographic tracers of structure with uncertain radial kernels is not guaranteed, and future work should quantify its performance on photometric clustering data -- the other key probe of the flagship ``3$\times$2pt'' analysis of imaging surveys -- and its cross correlation with cosmic shear and CMB lensing data \citep{K1K, DESY3, 2021JCAP...10..030G, Martin_white_tomo}. 

\section*{Acknowledgements}
  We would like to thank An\v e Slosar and Marius Millea for useful discussions. DA is supported by the Science and Technology Facilities Council through an Ernest Rutherford Fellowship, grant reference ST/P004474. PGF, CGG and AM are supported by European Research Council Grant No:  693024 and the Beecroft Trust. JRZ is supported by an STFC doctoral studentship. We made extensive use of computational resources at the University of Oxford Department of Physics, funded by the John Fell Oxford University Press Research Fund. 

  We made extensive use of the {\tt numpy} \citep{oliphant2006guide, van2011numpy}, {\tt scipy} \citep{2020SciPy-NMeth}, {\tt astropy} \citep{1307.6212, 1801.02634}, {\tt healpy} \citep{Zonca2019}, {\tt GetDist} \cite{2019arXiv191013970L}, and {\tt matplotlib} \citep{Hunter:2007} python packages. We also make use of the \texttt{Julia} packages {\tt ForwardDiff.jl} \citep{ForwardDiff} and {\tt Turing.jl} \citep{ge2018t}.

\section*{Data Availability}
  The code developed for this work as well as the derived datasets produced (power spectra and covariances) are available upon request. The catalogues and maps used were made publicly available by the authors of the relevant papers, as described in the text.

\bibliographystyle{mnras}
\bibliography{main,non_ads}

\appendix

\section{Stress-testing the approximation}\label{app:approx}

\begin{figure*}
  \centering
  \includegraphics[width=0.95\textwidth]{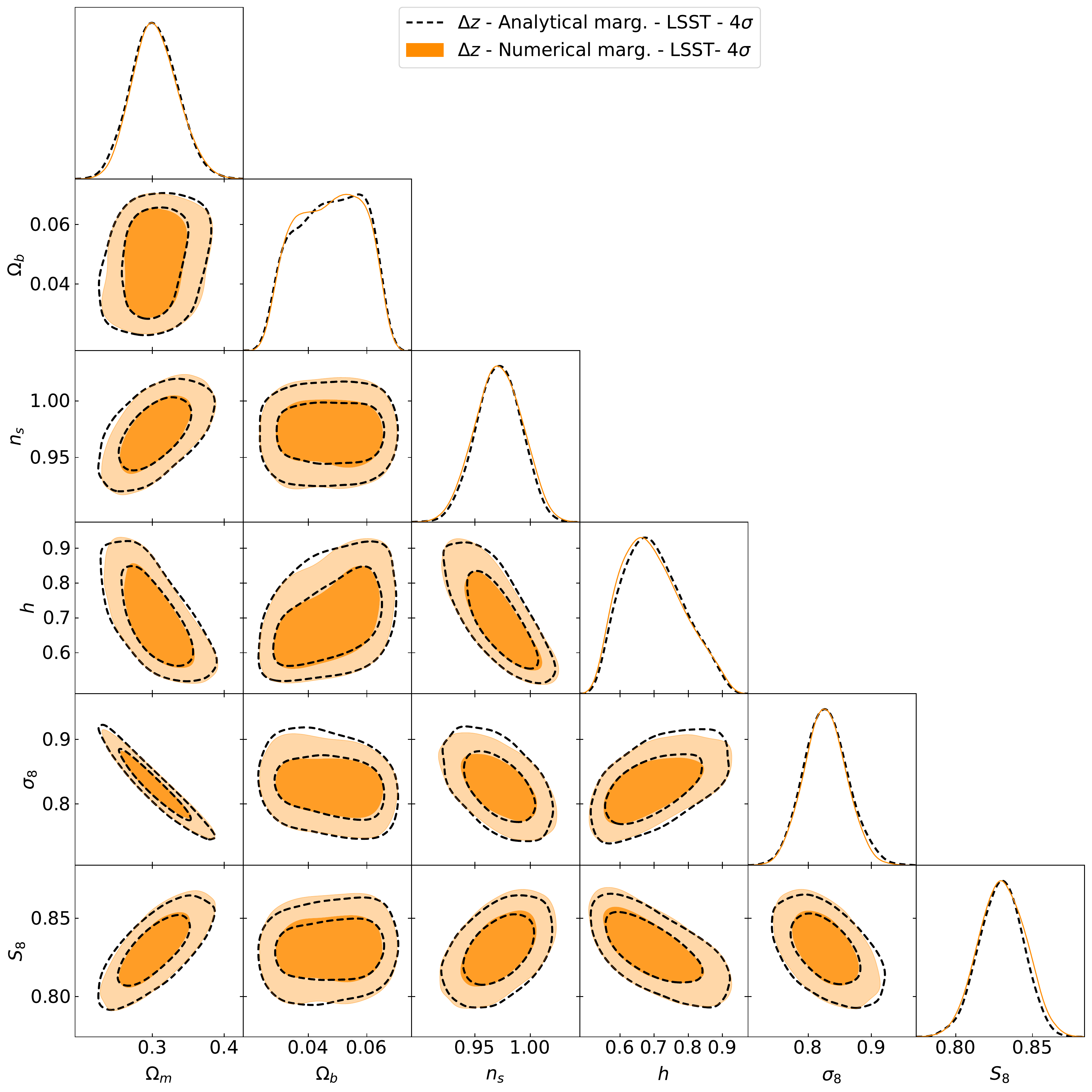}
  \caption{Shows a comparison between the obtained marginalised posterior distributions of the cosmological parameters when analytically marginalizing over the $\Delta z$ (black dashed) and when performing the full numerical marginalisation (orange) when analyzing LSST-like data. In both cases the $\Delta z$ prior distributions where made 4 times wider. We can observe that despite significantly broadening the prior distributions the analytical marginalisation returns virtually identical posteriors for the cosmological parameters.}
  \label{fig:4x}
\end{figure*}
  As described in Sect. \ref{sec:methods}, the approximation used here to analytically marginalise over the redshift calibration parameters assumes a sufficiently tight prior on these parameters, such that the dependence of the theory prediction on them can be linearised. Testing whether this assumption might break in a realistic scenario, is therefore essential. This is important in the context of Stage-IV since, even though it is expected that spectroscopic samples and the associated calibration techniques will improve over time, the increase in depth that LSST-like surveys will represent may make the calibration of the faintest samples in the survey particularly challenging.
  
  To further stress-test our approximate method, we repeat our analysis of the LSST-like futuristic data using the $\Delta z$ model for redshift uncertainties with priors 4 times larger than used in our fiducial analysis (which themselves were based on existing calibration samples). The result of this test is shown in Fig. \ref{fig:4x}. Reassuringly, the results show that, despite quadrupling the uncertainty in the redshift nuisance parameters, the analytic marginalisation method  yields virtually the same constraints on the cosmological parameters as the brute-force marginalisation, in spite of the significantly broader posterior contours. This implicit validates the approximation that a first-order expansion of the theory data vector with respect to a change in redshift distribution is sufficient over a conservative range of calibration priors.

\bsp
\label{lastpage}
\end{document}